\documentclass[prl, superscriptaddress, twocolumn, showpacs, floatfix, nobalancelastpage]{revtex4}

\usepackage{amsfonts}
\usepackage{amssymb}
\usepackage[dvips]{graphicx}
\usepackage{color}
\usepackage{amsmath}
\newcommand{\eend}      {\hspace{\stretch{1}}\rule{1ex}{1ex}}

\begin{document}

\title{Causality in Condensates: Grey Solitons as Remnants of BEC Formation.}

%\date{\today}
\author{Wojciech H. Zurek}
\affiliation{Theory Division, LANL, MS-B213, Los Alamos, NM  87545, USA}

\begin{abstract}
Symmetry breaking during phase transitions can lead to the formation of topological defects 
(such as vortex lines in superfluids). However, the usually studied BEC's 
%in elongated traps where superfluid has 
have the shape of a cigar,
%, its width comparable to the size of vortex core. 
a geometry that
impedes vortex formation, survival, and detection. I show that, in elongated traps, one can expect the formation of {\it grey solitons} (long-lived, non-topological ``phase defects'') as a result of the same 
%Kibble-Zurek %\cite{Kib76,Zur85a}
mechanism. Their number will rise approximately in proportion to the transition rate. This steep rise 
%(defect density usually increases with a fractional power of quench rate) 
is due to the increasing size of the region of the BEC cigar where the phase of the condensate 
wavefunction is chosen locally (rather than passed on from the already formed BEC).
%We extend the theory of symmetry breaking dynamics in non-equilibrium second order phase transitions known as the Kibble-Zurek mechanism (KZM) to transitions where the change of phase occurs not in time, but in space. This can be due to a time-independent spatial variation of a field that imposes a phase with one symmetry to the left of where it attains critical value, while allowing spontaneous symmetry breaking to the right of that critical borderline. Topological defects need not  form in such a situation. We show, however, that the size, in space, of the ``scar'' over which order parameter adjusts as it  ``bends'' interpolating between the phases with different symmetry follows from a KZM - like approach. 
%In particular, when the controlling external parameter changes on a scale $\lambda_Q$, and healing length far away from the critical point is given by $\xi_0$, the size of the region over which order parameter adjusts to its broken symmetry value is given by $\hat x = \sqrt {\lambda_Q \xi_0}$ in mean field theories. Similar formulae obtain for other universality classes or when the parameter change is not linear in the distance from its critical borderline. As we show on the example of a transverse quantum Ising, in quantum phase transitions this spatial scale -- size of the scar -- is directly reflected in the energy spectrum of the system: In particular, it determines the size of the energy gap.

\end{abstract}

\maketitle

%\section{Introduction}

Phase transitions are usually studied as equilibrium phenomena. However, as a consequence 
of the critical slowing down, second order phase transitions depart from equilibrium near the critical 
point, where the new broken symmetry phase is chosen. Hence, that choice must be made locally, 
within regions that can dynamically  ``agree'' on how to break symmetry.  Cosmology offers 
a well-known example: As pointed out by Kibble~\cite{Kib76}, relativistic causality alone limits the size 
of domains over which symmetry breaking can be coordinated. As a consequence, topological defects such as monopoles, cosmic strings, and domain walls can form.
% with significant cosmological consequences. 
%This is, for example, the source of the ``monopole problem'' that stimulated development of inflationary paradigm.

In laboratory phase transitions relativistic causality does not provide useful estimates of the domain size with the approximately uniform new phase, and, hence, does not lead to predictions of defect 
density. One can, however, estimate the domain size by appealing to universality of second order phase transitions~\cite{Zur85a}: Symmetry breaking is coordinated by the dynamics of the order 
parameter. In the vicinity of second order transitions critical slowing down implies that the relaxation 
time (which determines the reflexes of the system) and the healing length (which sets the scale on which its order parameter ``heals'', i.e., returns to its equilibrium value) diverge as:
\begin{eqnarray}
&&\tau={\tau_0} / {|\epsilon|^{\nu z}} 
\label{eq:tau1}\\
&&\xi= {\xi_0} / {|\epsilon|^{\nu}} 
\label{eq:xinu}
\end{eqnarray}
Above, ${\tau_0}$ and $ {\xi_0}$ depend on the microphysics, while the critical exponents $\nu$ and $z$ 
%are the critical exponents which 
define the universality class of the transition, and $\epsilon$ is the relative temperature
\begin{equation}
\epsilon=\frac {T_C-T} {T_C} \ ,
%; \ \ \ \ \epsilon=\frac {g-g_C} {g_C}
\label{eq:epsilon1}
\end{equation}
with $T_C$ the critical temperature. 

Taking the ratio of $\xi$ and $\tau$ one obtains
speed of sound:
\begin{equation}
v= (\xi_0 / \tau_0) |\epsilon|^{-(\nu-\nu z)} = v_0 |\epsilon|^{\nu(z-1)}  \ .
\label{eq:velsound}
\end{equation}
This is the speed of perturbations of the order parameter. The resulting sonic horizon plays a key role. 

Divergence of the healing length was recently observed in measurements of phase coherence {\it above} 
the BEC critical point~\cite{Ess}. In effect, the experiment of Esslinger et al. demonstrated that the phase 
of the condensate wavefunction is becoming coherent over distances that increase as the critical point 
is approached from above, as expected from Eq.~(\ref{eq:xinu}). If the critical region was traversed
infinitesimally slowly, all of the newly created BEC would have a single coherent phase. However,
when the transition is accomplished at a finite rate, critical slowing down, Eq.~(\ref{eq:tau1}), intervenes:
As its reflexes deteriorate, the phase of the order parameter cannot establish 
coherence over scales larger than the sonic horizon. 

In the usual discussions of topological defect formation \cite{Zur85a, Zur96a} one first calculates the instant $\hat t$ at which 
the system ceases to follow the externally imposed variation of its parameters by comparing the timescale 
$\epsilon/{\dot \epsilon}$ at which relative temperature changes to the relaxation time: 
\begin{equation}
\tau(\hat t) =  {\epsilon(\hat t)} / {\dot \epsilon(\hat t)} \  . 
\label{eq:that1}
\end{equation}
To obtain $\hat t$ we need the dependence of $\epsilon$ on $t$. We assume that it is linear, 
parametrized by quench time $\tau_Q$,
\begin{equation}
\epsilon= t / {\tau_Q} \ .
\label{eq:epst}
\end{equation}
The system adjusts its state adiabatically as long as the imposed rate of change is slow compared 
to its reflexes given by the inverse of $\tau$, Eq.~(\ref{eq:tau1}). The transition
from the adiabatic to impulse behavior happens at $\hat t$ given by
Eq.~(\ref{eq:that1}), i.e.
%\begin{equation}
${\tau_0} {|\frac {\hat t}{\tau_Q}|^{-\nu z}}=\hat t \ .$
%\label{eq:that2}
%\end{equation}
So, the order parameter  ``freezes'' when the relaxation time and $t$ coincide;
\begin{equation}
\hat t = (\tau_0\tau_Q^{\nu z})^{\frac 1 {1+\nu z}}=\hat \tau \ .
\label{eq:that3}
\end{equation}
%Here and below we use caret (``$\hat " "$") to mark various quantities computed at this freezeout instant.

The order parameter
%(with its phases ``cohered'' over distances reported in Ref.~\cite{Ess})
will resume evolution only $\hat t$ after critical point is passed. The scale of the fluctuations 
(reported in Ref.~\cite{Ess}) that seed structures (such as topological defects) in the broken 
symmetry BEC phase \cite{Zur85a, Zur96a,  Kib03, Kib07} is thus established at $\hat t$, i.e., at the relative temperature:
\begin{equation}
\hat \epsilon = \bigl(\frac {\tau_0} {\tau_Q} \bigr)^{\frac 1 {1+\nu z}}
\label{eq:epshat}
\end{equation}
The scale given by the corresponding healing length
\begin{equation}
\hat \xi = \xi_0\bigl(\frac {\tau_Q} {\tau_0} \bigr)^{\frac \nu {1+\nu z}}
\label{eq:xihat}
\end{equation}
determines the density of defects. The phase of the newly formed BEC wavefunction will be coherent on scales $\sim \hat \xi$. Therefore, one expects a defect fragment (e.g., one section of a vortex line) per 
$\hat \xi$-sized domain~\cite{Kib76}. 
%In course of BEC formation
% and in accord with Refs.~\cite{Zur85a, Zur96a} -- supported by measurements of phase coherence above BEC transition, Ref. \cite{Ess} -- 
%
In a homogeneous 3D quench 
this leads to vortex line density of $\sim\hat \xi^{-2}$~\cite{Zur85a, Zur96a,AZ99a}, in accord
with most  of the experimental evidence~\cite{Kib07}, including  
BEC's~\cite{And07a,And07b}. It is confirmed and refined by numerics~\cite{LZ96a}, which 
also indicate that there is typically less than one defect fragment per $\hat \xi$-sized domain: Rather, a defect fragment in $f\hat \xi$-sized region, where $\tau_Q$-independent $f$ is set by microphysics of the transition, is typical. The factor $f$ can be greater than 1, 
and $f\sim10$ are common~\cite{LZ96a}. The density of defects created by phase transitions is 
the best known (but not the only) prediction of this ``Kibble-Zurek mechanism'' (or ``KZM''). 
%This prediction exploits equilibrium properties of the system -- scaling behavior in the vicinity of the critical point -- to predict non-equilibrium consequences of the quench. 
%KZM was used (see e.g. Ref. \cite{And07a}) to deduce density of vortices in bulk BEC. 

In the inhomogeneous case (e.g., effectively 1D trap) situation is different: The gas density (and, hence, local critical temperature $T_C$) depends on location. Thus, even when $T$ drops uniformly due to evaporative cooling, the gas will reach local critical temperature $T_C(\vec r)$ at different instants: $\epsilon(\vec r_F,t_F)=0$ defines the {\it front of the transition} $\vec r_F(t_F)$ as it spreads through the cigar. So the critical front will appear at $t_F$ that depends on location $\vec r_F$. 

Before the evaporative cooling the local density is~\cite{Bagnato}:
\begin{equation}
\rho(x) = \rho_0 \exp(-\beta V (\vec r)) \ .
\label{eq:density}
\end{equation}
Above $V(\vec r)$ is (typically, harmonic) trap potential and $\beta=1/k_BT$. Einstein's condition for 
BEC formation involves density and de Broglie wavelength, $\rho\lambda_{dB}^3(T_C)\approx 2.61$. 
In elongated traps one can in effect eliminate transverse dimensions  \cite{Ketterle}. 
This implies a {\it local} $T_C(x)$:
\begin{equation}
T_C(x)\simeq \frac {2 \pi \hbar^2} {m k_B} \left(\frac {\rho(x)} {2.61}\right)^{2/3} \ ,
\label{eq:TC}
\end{equation}
where $m$ is the mass of bosons, while $\hbar$ and $k_B$ are Planck and Boltzmann constants.
In other words, when anywhere in a large effectively 1D harmonic trap temperature falls below local $T_C(x)$ 
in a region large compared to the healing length,
the condensate will begin to form.

%The critical temperature will be therefore reached at different points $\vec r_F$ at different instants $t_F$.
We assume that cooling decreases $T$ uniformly, so that;
\begin{equation}
T(t) = T_C(0)\left( 1- \frac t {\tau_Q}\right)
\label{eq:T(t)}
\end{equation}
everywhere in the trap. Therefore, front coordinates $x_F$ and $t_F$ are related by the equation 
$\epsilon(x_F,t_F)=0$, or:
\begin{equation}
\frac {t_F} {\tau_Q} =1- \frac {T_C(x_F)} { T_C(0)} 
\label{eq:t_F}
\end{equation}

So the condensate can form first in a healing length size domain near $x=0$, where the potential is deepest. In an infinitesimally slow quench that initial seed would grow to occupy the whole trap. But 
this cannot happen when quench is so fast that regions far away from the center quickly attain 
temperatures far below the local $T_C(x)$: They will begin to form BEC independently, from local 
seeds, and with locally selected phases. 

The phase of the newly formed BEC wavefunction can be then either communicated along $x$, or selected at 
different points of the trap independently (as would be the case in a homogeneous quench). What 
actually happens is decided by causality, and depends on the comparison of the causal
horizon defined by the relevant sound velocity, Eq.~(\ref{eq:velsound}), and the velocity of the front:
\begin{equation}
v_F= \left|\frac {d x_F} {d t_F}\right| = \frac {T_C(0)} {\tau_Q} \left| \frac {d T_C(x)}{d x_F} \right|^{-1}
 \label{eq:v_F}
\end{equation}
The speed of the front is infinite at the center of the trap where $V(x)$ has its minimum, and drops
with the inverse of the gradient of the critical temperature. 
The perturbations travel distance $\sim\hat \xi$ 
over time $\hat t$. So, the relevant speed of sound corresponds to the freezeout $\hat \epsilon$:
\begin{equation}
\hat v =  \frac {\hat \xi} {\hat \tau} = \frac {\xi_0} {\tau_0} \left(\frac {\tau_0} {\tau_Q} \right)^{\frac {\nu(z-1)} {1+\nu z}}.
\label{eq:vhat}
\end{equation}
The role of $\hat v$ and its sonic horizon emerged in discussions of vortex formation in $^3$He superfluid. These experiments start with a cigar-shaped bubble heated above the critical 
point~\cite{He3} which quickly cools to the temperature of the surrounding  $^3$He superfluid. One might have expected that superfluid on the outside of the bubble will impose (uniform) phase of its
wavefunction on the cooling ``cigar''. That this need not happen was noted in Ref.~\cite{KV}: 
When the front $T(x)=T_C(x)$ spreads faster than $\hat v$, the phase of the newly formed condensate 
is chosen locally. Subsequent studies~\cite{DLZ} confirmed that when the front velocity $v_F$ exceeds 
$\hat v$, symmetry breaking happens as in a homogeneous transition, and defects appear with density inferred from $\hat \xi$. However, when $\hat v > v_F$, preexisting condensate propagates its 
phase into the newly forming regions, and topological defects do not from.

In the quasi-1D traps one does not expect to see vortices as, at formation, their $\hat \xi$-sized cores barely fit inside the cigar, so, even if they form, they can easily escape. Vortices do form in 
quasi-2D pancake traps~\cite{And07a,And07b}. But in the effectively 1D geometry there is a stable defect related to phase nonuniformity -- the grey soliton~\cite{PitStri}. It corresponds to a solution of the Gross-Pitaevski equation, and describes a localized (healing length scale) nonuniformity of BEC phase, and a corresponding depletion of condensate density. The solitons are not 
topological:  The phase change across the soliton can be arbitrary but, far ($\sim \xi$) away, it asymptotes to a constant value. Its change by $\pi$ yields a {\it dark} soliton, which causes complete depletion of the BEC density at its center. Dark solitons are stationary, 
but {\it grey} solitons, with phase change less then $\pi$, and a smaller depletion 
of central density (hence ``grey" in their name), move along BEC cigar with velocities set by the local 
density depletion. When they arrive at the point where BEC density is lower, they become 
(locally) ``dark",  stop, and are reflected. Grey solitons were seen oscillating in this manner along 
BEC cigars~\cite{Den}.

We can expect that non-uniformities of phase left by the BEC formation will give rise to grey solitons. 
Using KZM we can estimate density of phase jumps caused by the BEC formation. Thus, we can also 
estimate density of grey solitons in a homogeneous region, and their total number left in the trap 
by the phase transition into BEC. 

To this end, we compute local ``freezeout'' values of $\hat \xi$, $\hat \tau$, and $\hat v$ using the local rate 
of change of $\epsilon$:
\begin{equation}
\frac {d \epsilon (x,t)} {d t}|_x \ = \ \frac {T_C(0)}{T_C(x)} \frac 1 {\tau_Q} = \frac 1 {\tau_Q(x)}
\label{eq:tauQ(x)}
\end{equation}
This defines effective local quench time 
\begin{equation}
\tau_Q (x)=\tau_Q~\frac{T_C(x)}{T_C(0)}, 
\label{eq:tauQ1(x)}
\end{equation}
which in turn yields local relative temperature:
\begin{equation}
\epsilon (x,t) = \frac {t-t_F(x)} {\tau_Q(x)}
\label{eq:epsilonx}
\end{equation}
Now one can proceed as usual and compute local $\hat t$:
\begin{equation}
\hat t_x=\left(\tau_0 (\tau_Q \frac {T_C(x)} {T_C(0)})^{\nu z} \right)^{\frac 1 {1+\nu z}} = \left( \tau_0 \tau_Q^{\nu z}(x) \right)^{\frac 1 {1+\nu z} }
\label{eq:hatt_x}
\end{equation}
Note that this $\hat t_x$ gives the time interval to the instant $t_F(x)$ at which the critical point is
reached at the location $x$, and Eq.~(\ref{eq:tauQ(x)}) is satisfied. This corresponds to the local
\begin{equation}
\hat \epsilon_x = \frac {\hat t_x} {\tau_Q(x)} = \bigl(\frac {\tau_0} {\tau_Q(x)}\bigr)^{\frac 1 {1+\nu z} }
\label{eq:epsx}
\end{equation}

We have now all the ingredients to calculate the local frozen out healing length:
\begin{equation}
\hat \xi_x = \frac {\xi_0} {{\hat \epsilon_x}^\nu} = \xi_0 \bigl(\frac {\tau_Q(x)} {\tau_0}\bigr)^{\frac \nu {1+\nu z} }
\label{eq:hatxix}
\end{equation}
Local velocity at the freezeout is then a function of $x$:
\begin{equation}
\hat v_x = \frac {\hat \xi_x} {\hat \tau(x)} = \frac {\xi_0} {\tau_0} \bigl(\frac {\tau_0} {\tau_Q(x)}\bigr)^{\frac {\nu(z-1)} {1+\nu z} }
\label{eq:vx}
\end{equation}
These estimates are essentially the same as for the homogeneous case: Key modification enters
through the locally defined $\tau_Q(x)$, Eq.~(\ref{eq:tauQ1(x)}). 

\begin{figure}[tp]
\centering  
\vspace{-0.1in}
\includegraphics[width=10cm]{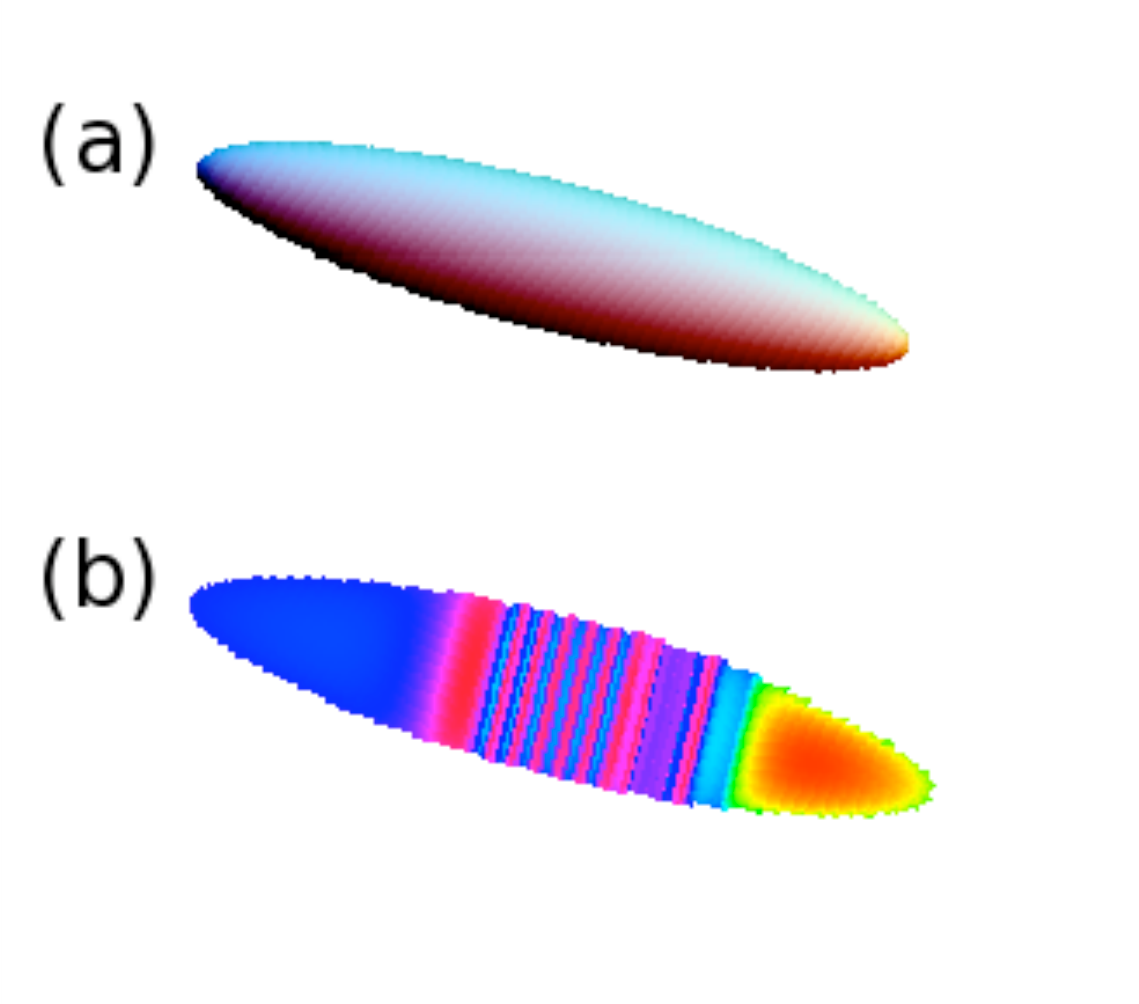}
\vspace{-0.5in}
\caption{{\it Formation of grey solitons in a cigar-shaped Bose-Einstein Condensate.}
{\bf (a)} Isodensity contour in the trapped gas. As evaporative cooling proceeds, critical 
temperature is first reached in the center of the trap. That is where the condensate will form first. 
When cooling is sufficiently slow, this initial seed grows and imposes its selection of the condensate wavefunction phase on the whole cigar, and no grey solitons are created by the quench. 
{\bf (b)} As is seen in the schematic color plot of the wavefunction phase in a cross section of a BEC cigar, the situation changes when BEC phase front -- the location where the decreasing temperature 
is instantaneously equal to the local critical temperature (set by the local density via Einsten's condition)
-- moves faster than the velocity $\hat v$ with which perturbations of the emerging order parameter can spread. In this case regions of size $\hat \xi$, the relevant 
healing length, select phase of the BEC wavefunction independently. The front velocity $v_F$ 
is infinite at $x=0$, but falls rapidly with the distance $x$ from the center. Condensate phase will be selected randomly by the symmetry breaking process in regions where $\hat v < v_F$. 
Such random phase distribution provides seeds for grey solitons. Phase front moves much less rapidly in the narrow direction of the cigar, so phases selected near the axis spread sideways, resulting in phase stripe pattern seen above in the schematic view of the BEC ``cigar''.}
\label{lzfkzm}
\end{figure}

These predictions apply to the central part of the cigar where the critical front spreads faster 
than $\hat v$ -- than the velocity of the perturbations of the order parameter. The region where 
the above quasi-homogeneous quench predictions are accurate must therefore satisfy $v_F > \hat v_x$. 
In view of our above discussion this leads to:
\begin{equation}
\frac {T_C(0)} {\tau_Q} \left| \frac { dT_C(x) } {d x}\right|^{-1} > \frac {\xi_0} {\tau_0} \left(\frac {\tau_0} {\tau_Q(x)}\right)^{\frac {\nu(z-1)} {1+\nu z} }
\label{eq:KV}
\end{equation}
When $V(x) = \frac {m\omega^2x^2}2$, $T_C(x)$, Eq.~(\ref{eq:TC}), is a Gaussian:
\begin{equation}
T_C(x)=T_C(0) e^{-x^2 / 2\Delta^2}
\label{eq:Gauss}
\end{equation}
where $\Delta^{-2}=\frac 2 3 \beta m \omega^2$, and we ignored variations perpenicular to the long axis. Inequality (\ref{eq:KV}) leads to:
\begin{equation}
|\hat X| < \frac {\Delta^2} {\xi_0} \bigl(\frac {\tau_0} {\tau_Q}\bigr)^{\frac {1+\nu} {1+\nu z} } 
e^{ 
{\frac {(1+\nu){\hat X}^2} {{2(1+\nu z)}{\Delta^2}} }} 
\label{eq:hatX}
\end{equation}
This inequality determines size of the section $[-\hat X, \hat X]$ of the cigar where $v_F>\hat v$,
and the motion of the critical point is supersonic. There the quench is effectively homogeneous, and 
defects (including solitons) will appear with separations given by the local $\hat \xi$ (see Fig. 1).

The equation for $\hat X$ is simple, but it is transcendental. We focus on the case where $\hat X < \Delta$.
Then the exponent in Eq.~(\ref{eq:hatX}) can be expanded, which leads to: 
%the following simple estimate of $\hat X$:
\begin{equation}
|\hat X| \approx \frac {\Delta^2} {\xi_0} \bigl(\frac {\tau_0} {\tau_Q}\bigr)^{\frac {1+\nu} {1+\nu z} } 
\label{eq:hatXapprox}
\end{equation}
%Our discussion above leads us to believe that 
This estimate of $\hat X$ is valid for slow quenches, i.e. it breaks down when $ \frac {\Delta} {\xi_0} \bigl(\frac {\tau_0} {\tau_Q}\bigr)^{\frac {1+\nu} {1+\nu z} } > 1$, 
but holds when:
\begin{equation}
\tau_Q \geq \tau_0  \bigl(\frac {\Delta} {\xi_0} \bigr)^{\frac {1+\nu z} {1+\nu } } = \tau_0  \bigl(\frac {\Delta} {\lambda_{dB}} \bigr)^{\frac {1+\nu z} {1+\nu } }
\label{eq:hatX1approx}
\end{equation}
We assume that this is indeed the case. This focus on slow quenches is anyway prudent: Our discussion assumes that, outside of the freezeout interval, order parameter is at or near the equilibrium set by the relative temperature $\epsilon(t)$. Very rapid quenches could strain this assumption (although it is unlikely they could be implemented using evaporative cooling, as collisions that control its rate 
also assure evolution of the order parameter). 

Equation (\ref{eq:hatX1approx}) yields simple scaling for the total
number of solitons. Note that above we have set ${\xi_0}={\lambda_{dB}}$, de Broglie 
wavelength at the critical temperature~\cite{AZ99a}.  
We are now ready to estimate the total number of grey solitons. We 
obtain it by multiplying the size of the quasi-homogeneous quench region by the expected density 
of phase changes. This yields:
\begin{equation}
N \approx \frac {2 \hat X} {f \hat \xi} = \frac { 2 \Delta^2} {f \lambda_{dB}^2} \bigl( \frac {\tau_0} {\tau_Q} \bigr)^{\frac {1+2\nu} {1+\nu z} } 
\label{eq:Ngrey}
\end{equation}
The surprise is that scaling of the number of solitons with the quench timescale $\tau_Q$ is so steep. 
For example, for the plausible values $\nu= \frac 2 3 $ and $z= \frac 3 2$ we predict $\frac {1+2\nu} {1+\nu z} = \frac 7 6$, while for mean field $\nu= \frac 1 2 $ and $z= 2$ the exponent ${\frac {1+2\nu} {1+\nu z} }=1$. So the number of grey solitons is expected to be approximately proportional to the quench rate.

Several aspects of the above prediction deserve comment. To begin, note that we have 
ignored all the aspects of the process that cannot be deduced from the universality class. They will
influence size of $f$. Here we include issues such as how dark a grey soliton must be to count as
 a soliton, and other matters relevant for experiments. For instance, it is known that solitons
 -- while they are long-lived -- do not live forever. Therefore, the number of solitons will depend on
 their survival rates. 
 
 The calculation above also addresses the question of when the quench can produce a uniform BEC. 
 This will happen when $\hat \xi \gg \hat X$, for quenches so slow that they produce $N\ll 1$ 
 solitons in a trap. There is also an opposite limit of very fast quenches. We shall not address it here 
 as it is cumbersome (e.g., transcendental Eq.~(\ref{eq:hatX}) cannot be approximated in a way that
 yields a simple result). Moreover, in order to reach it in experiments one would need to drop
 temperature very quickly throughout the trap, and far below $T_C$ at the center of the trap. For
 such rapid quenches linear approximation $\epsilon= t / {\tau_Q} $ is likely to break down, leading 
 to further cumbersome but trivial complications. This last comment brings one more remark: 
 In most experimental settings $T(t)$ will 
fall below $T_C$ at the center of the trap, but this may be above $T_C(x)$
 at some sufficiently large $x$. Our analysis applies as long as that $x$ lies outside the central
 interval of $2\hat X$, or, more precisely, as long as the quench can be well approximated by 
 linear relations (e.g., Eq.~(\ref{eq:T(t)}) inside it.
 
We discussed formation of grey solitons in elongated traps. The experiment aimed at detecting such non-topological remnants of a BEC phase transition should be easier than experiments~\cite{And07b} that study formation of vortex lines (which require more that a quasi-1D geometry). It should
allow one to probe the connection between causality and symmetry breaking and test
scalings predicted by the Kibble-Zurek mechanism.
%Indeed, such experiment is being undertaken by Peter Engels ~\cite{Engels}.
 
 I would like to thank Malcolm Boshier, Bogdan Damski, and especially Peter Engels for stimulating discussions. This research
 was supported by DoE under the LDRD program at Los Alamos.

\end{document}